\pdfoutput=1
\documentclass[prb,twocolumn,showpacs,aps]{revtex4}
\usepackage{graphicx,graphics,color,epsfig}% Include figure files
\usepackage{bm}
\usepackage{amsmath}
\usepackage{amssymb}
\usepackage{epstopdf}

\makeatletter

\newcommand{\Rmnum}[1]{\expandafter\@slowromancap\romannumeral #1@}
\makeatother

\usepackage{color}

\begin{document}

\title{Probing active/passive bands by quasiparticle interference in Sr$_{2}$RuO$_{4}$}

\author{Yi Gao,$^{1}$ Tao Zhou,$^{2}$ Huaixiang Huang,$^{3}$ C. S. Ting,$^{4}$ Peiqing Tong,$^{1}$ and Qiang-Hua Wang $^{5}$}
\affiliation{$^{1}$Department of Physics and Institute of Theoretical Physics,
Nanjing Normal University, Nanjing, 210023, China\\
$^{2}$College of Science, Nanjing University of Aeronautics and
Astronautics, Nanjing, 210016, China\\
$^{3}$Department of Physics, Shanghai University, Shanghai, 200444, China\\
$^{4}$Department of Physics and Texas Center for Superconductivity, University of Houston, Houston, Texas, 77204, USA\\
$^{5}$National Laboratory of Solid State Microstructures, Nanjing University, Nanjing, 210093, China}

\begin{abstract}
The quasiparticle interference (QPI) in Sr$_{2}$RuO$_{4}$ is theoretically studied based on two different pairing models in order to propose an experimental method to test them. For a recently proposed two-dimensional model with pairing primarily from the $\gamma$ band, we found clear QPI peaks evolving with energy and their locations can be determined from the tips of the constant-energy contour (CEC). On the other hand, for a former quasi-one-dimensional model with pairing on the $\alpha$ and $\beta$ bands, the QPI spectra are almost dispersionless and may involve off-shell contributions to the scatterings beyond the CEC. The different behaviors of the QPI in these two models may help to resolve the controversy of active/passive bands and whether Sr$_{2}$RuO$_{4}$ is a topological superconductor.

\end{abstract}


\maketitle

\emph{Introduction}.---Superconductivity was found in Sr$_{2}$RuO$_{4}$ by Maeno \emph{et al.} in 1994.~\cite{maeno} Soon after, it was proposed that the superconducting (SC) pairing symmetry in this kind of material may be $p$-wave.~\cite{rice,baskaran} Later experiments suggest that the SC state has odd parity \cite{nelson,duffy,ishida2} and spontaneously breaks time-reversal symmetry.~\cite{luke,kidwingira,xia} Thus Sr$_{2}$RuO$_{4}$ is a possible candidate for a chiral $p$-wave superconductor.~\cite{mackenzie,kallin} Recently the chiral $p$-wave superconductor has attracted much attention since it may give rise to a topological superconductor which supports gapless modes at the edge of the system and in vortex cores. Relaxing the weak spin-orbital coupling (e.g., by a suitably applied magnetic field) such zero modes may become Majorana modes. They are robust against perturbations, thus are proposed to be the building blocks for quantum computation.~\cite{read,nayak}

Up to now, whether Sr$_{2}$RuO$_{4}$ can be viewed as a topological superconductor is still controversial. For example, in Sr$_{2}$RuO$_{4}$, there are three energy bands cut by the Fermi level, denoted as $\alpha$, $\beta$ and $\gamma$.~\cite{oguchi,mackenzie2,shen} The $\alpha$ and $\beta$ bands are quasi-one-dimensional and are composed of the Ru $d_{xz}$ and $d_{yz}$ orbitals while the $\gamma$ band is two-dimensional and is from the Ru $d_{xy}$ orbital. The specific heat \cite{deguchi} and nuclear spin relaxation measurements,~\cite{Ishida} as well as early calculations \cite{Agterberg} suggest that superconductivity occurs only in a subset of the bands. Previous theories concluded that the SC pairing arises in the $\gamma$ band and is of chiral $p$-wave symmetry [$\mathbf{d}(\mathbf{k})=\Delta_{0}(\sin k_{x}+i\sin k_{y})\mathbf{z}$].~\cite{miyake,ng} In this case Sr$_{2}$RuO$_{4}$ is a topological superconductor. However, the predicted magnitude of edge current has not been detected experimentally.~\cite{kirtley1,kirtley2} This discrepancy leads Raghu \emph{et al.} to propose that,~\cite{raghu} instead of the $\gamma$ band, the SC pairing in Sr$_{2}$RuO$_{4}$ should take place in the $\alpha$ and $\beta$ bands (denoted as 1D model). In this case, the $d$ vector can be written as
\begin{eqnarray}
\label{draghu}
\mathbf{d}_{1}(\mathbf{k})&=&\Delta_{0}\sin k_{x}\cos k_{y}\mathbf{z},\nonumber
\end{eqnarray}
\begin{eqnarray}
\mathbf{d}_{2}(\mathbf{k})&=&i\Delta_{0}\sin k_{y}\cos k_{x}\mathbf{z},
\end{eqnarray}
where $1$ and $2$ refer to the $d_{xz}$ and $d_{yz}$ orbitals, respectively. Although the pairing is still chiral, it is not a topological superconductor since the skyrmion numbers on the hole-like $\alpha$ and electron-like $\beta$ bands cancel out, thus explaining the absence of edge current. Recently Wang \emph{et al.} performed a comprehensive functional renormalization group study of the pairing mechanism in Sr$_{2}$RuO$_{4}$ considering all of the three bands on equal footing.~\cite{wang} Their conclusions are: Superconductivity arises primarily in the $\gamma$ band (denoted as 2D model) where the $d$ vector can be approximated as
\begin{eqnarray}
\label{d}
\mathbf{d}(\mathbf{k})&\sim&[p_{1}\sin k_{x}+p_{2}\cos k_{y}\sin k_{x}\nonumber\\
&&+i(p_{1}\sin k_{y}+p_{2}\cos k_{x}\sin k_{y})]\mathbf{z},
\end{eqnarray}
with $p_{1}/p_{2}\sim -0.4375$. Although it is a topological superconductor, there are deep gap minima which make the edge current fragile against small perturbations. While this does not rule out the edge current, it nonetheless reconciles the difficulty in the experimental detection. The theory also predicts that the pairing on the $(\alpha,\beta)$ bands are even smaller than the minimum on the $\gamma$ band.

In this paper, we propose to measure the quasiparticle interference (QPI)~\cite{QPI} as a method to resolve the above controversy. The idea is, in realistic materials, an incoming wave is scattered into an outgoing wave by some elastic impurities and the interference between these two waves gives rise to a spatial modulation of the local density of states (LDOS), which can be measured by scanning tunneling microscopy (STM).~\cite{hoffman} By inspecting the modulation wave vectors, the information of the electronic band structure as well as the SC pairing can be obtained. For Sr$_{2}$RuO$_{4}$, if the SC pairing occurs in the $\gamma$ ($\alpha$ and $\beta$) band, then the modulation of the LDOS due to that band will change once superconductivity sets in, thus changing the scattering wave vector $\mathbf{q}$ of that band. Therefore by comparing the QPI in the normal and SC states, we can gain information on the active band(s) in which superconductivity develops.

\emph{Method}.---We start with the lattice models proposed in Refs.~\onlinecite{raghu} and~\onlinecite{wang}. The Hamiltonian can be written as
\begin{eqnarray}
\label{h}
H&=&\sum_{\mathbf{k}}\varphi_{\mathbf{k}}^{\dag}M_{\mathbf{k}}\varphi_{\mathbf{k}},\nonumber\\
\varphi_{\mathbf{k}}^{\dag}&=&(c_{\mathbf{k}1\uparrow}^{\dag},c_{\mathbf{k}2\uparrow}^{\dag},c_{\mathbf{k}3\uparrow}^{\dag},c_{-\mathbf{k}1\downarrow},c_{-\mathbf{k}2\downarrow},c_{-\mathbf{k}3\downarrow}),\nonumber\\
M_{\mathbf{k}}&=&\begin{pmatrix}
H_{\mathbf{k}}&\Delta_{\mathbf{k}}\\\Delta_{\mathbf{k}}^{\dag}&-H_{-\mathbf{k}}
\end{pmatrix},\nonumber\\
H_{\mathbf{k}}&=&\begin{pmatrix}
\epsilon_{1\mathbf{k}}&\epsilon_{12\mathbf{k}}&0\\\epsilon^{*}_{12\mathbf{k}}&\epsilon_{2\mathbf{k}}&0\\0&0&\epsilon_{3\mathbf{k}}
\end{pmatrix},
\end{eqnarray}
where $c_{\mathbf{k}1\uparrow}^{\dag}$, $c_{\mathbf{k}2\uparrow}^{\dag}$ and $c_{\mathbf{k}3\uparrow}^{\dag}$ create a spin $\uparrow$ electron with momentum $\mathbf{k}$ in the $d_{xz}$, $d_{yz}$ and $d_{xy}$ orbitals, respectively and $\Delta_{\mathbf{k}}$ is a $3\times3$ matrix. For the 2D model, we have
\begin{eqnarray}
\label{twod}
\epsilon_{1\mathbf{k}}&=&-2t_{1}\cos k_{x}-t_{0},\nonumber\\
\epsilon_{2\mathbf{k}}&=&-2t_{1}\cos k_{y}-t_{0},\nonumber\\
\epsilon_{12\mathbf{k}}&=&-4t_{2}\sin k_{x}\sin k_{y},\nonumber\\
\epsilon_{3\mathbf{k}}&=&-2t_{3}(\cos k_{x}+\cos k_{y})-4t_{4}\cos k_{x}\cos k_{y}+t_{5}-t_{0},\nonumber\\
\Delta_{\mathbf{k}}^{33}&=&\Delta_{0}[p_{1}\sin k_{x}+p_{2}\cos k_{y}\sin k_{x}\nonumber\\
&&+i(p_{1}\sin k_{y}+p_{2}\cos k_{x}\sin k_{y})].
\end{eqnarray}
Here $t_{0-5}=1.1,1,0.1,0.8,0.35,-0.2$, $p_{1}=-0.4375$ and $p_{2}=1$. In this model we ignore the tiny pairing on the $(\alpha,\beta)$ bands henceforth. This approximation does not alter the conclusion as long as the quasiparticle energy is above the related small energy scale. For the 1D model, we have
\begin{eqnarray}
\label{oned}
\epsilon_{1\mathbf{k}}&=&-2t_{1}\cos k_{x}-2t_{2}\cos k_{y}-t_{0},\nonumber\\
\epsilon_{2\mathbf{k}}&=&-2t_{1}\cos k_{y}-2t_{2}\cos k_{x}-t_{0},\nonumber\\
\epsilon_{12\mathbf{k}}&=&-2t_{5}\sin k_{x}\sin k_{y}+it_{5},\nonumber\\
\epsilon_{3\mathbf{k}}&=&-2t_{3}(\cos k_{x}+\cos k_{y})-4t_{4}\cos k_{x}\cos k_{y}-t_{0},\nonumber\\
\Delta_{\mathbf{k}}^{11}&=&\Delta_{0}\sin k_{x}\cos k_{y},\nonumber\\
\Delta_{\mathbf{k}}^{22}&=&i\Delta_{0}\sin k_{y}\cos k_{x},
\end{eqnarray}
with $t_{0-5}=1,1,0.1,0.8,0.3,0.1$. Here the $it_{5}$ term in $\epsilon_{12\mathbf{k}}$ is from the spin-orbital coupling. In both the models, we set $\Delta_{0}=0.03$. The normal state is represented by setting $\Delta_{\mathbf{k}}=0$ in Eq. (\ref{h}). When a single impurity is located at the origin, the impurity Hamiltonian can be written as
\begin{eqnarray}
\label{himp}
H_{imp}&=&V_{s}\sum_{m=1}^{3}\sum_{\sigma=\uparrow,\downarrow}c_{0m\sigma}^{\dag}c_{0m\sigma}\nonumber\\
&=&\frac{V_{s}}{N}\sum_{m=1}^{3}\sum_{\sigma=\uparrow,\downarrow}\sum_{\mathbf{k},\mathbf{k}^{'}}c_{\mathbf{k}m\sigma}^{\dag}c_{\mathbf{k}^{'}m\sigma},
\end{eqnarray}
here $N$ is the system size ($396\times396$ in the following). We consider nonmagnetic impurity only, diagonal in the orbital basis and with a scattering strength $V_{s}$=4$\Delta_{0}$ for definiteness. Following the standard $T$-matrix procedure,~\cite{zhu} we define the Green's function matrix as
\begin{eqnarray}
\label{gt}
g(\mathbf{k},\mathbf{k}^{'},\tau)=-\langle T_{\tau}\varphi_{\mathbf{k}}(\tau)\varphi_{\mathbf{k}^{'}}^{\dag}(0)\rangle,
\end{eqnarray}
and
\begin{eqnarray}
\label{gw}
g^{R/A}(\mathbf{k},\mathbf{k}^{'},\omega)&=&\delta_{\mathbf{k}\mathbf{k}^{'}}g_{0}^{R/A}(\mathbf{k},\omega)\nonumber\\
&&+g_{0}^{R/A}(\mathbf{k},\omega)T^{R/A}(\omega)g_{0}^{R/A}(\mathbf{k}^{'},\omega).
\end{eqnarray}
Here $R$ and $A$ refer to the retarded and advanced Green's function, respectively and
\begin{eqnarray}
\label{g0}
g_{0}^{R/A}(\mathbf{k},\omega)&=&[(\omega\pm i0^{+})I-M_{\mathbf{k}}]^{-1},\nonumber\\
T^{R/A}(\omega)&=&[I-\frac{U}{N}\sum_{\mathbf{q}}g_{0}^{R/A}(\mathbf{q},\omega)]^{-1}\frac{U}{N},\nonumber\\
\end{eqnarray}
where $I$ is a $6\times6$ unit matrix and
\begin{eqnarray}
\label{u}
U^{nn}=\begin{cases}
V_{s}&\text{$n=1,2,3$},\\
-V_{s}&\text{$n=4,5,6$}.
\end{cases}
\end{eqnarray}
The experimentally measured LDOS is expressed as
\begin{eqnarray}
\label{rour}
\rho(\mathbf{r},\omega)&=&-\frac{1}{\pi}\sum_{m=1}^{3}\sum_{\sigma=\uparrow,\downarrow}{\rm Im}\langle\langle c_{\mathbf{r}m\sigma}|c_{\mathbf{r}m\sigma}^{\dag}\rangle\rangle_{\omega+i0^{+}}\nonumber\\
&=&-\frac{1}{\pi N}\sum_{m=1}^{3}\sum_{\mathbf{k},\mathbf{k}^{'}}{\rm Im}\Big{\{}[g_{mm}^{R}(\mathbf{k},\mathbf{k}^{'},\omega)\nonumber\\
&&-g_{m+3m+3}^{A}(\mathbf{k},\mathbf{k}^{'},-\omega)]e^{-i(\mathbf{k}-\mathbf{k}^{'})\cdot\mathbf{r}}\Big{\}},
\end{eqnarray}
and its Fourier transform is defined as $\rho(\mathbf{q},\omega)=\sum_{\mathbf{r}}\rho(\mathbf{r},\omega)e^{i\mathbf{q}\cdot\mathbf{r}}$, which can be expressed as
\begin{widetext}
\begin{eqnarray}
\label{rouq}
\rho(\mathbf{q},\omega)&=&-\frac{1}{2\pi}\sum_{m=1}^{3}\sum_{\mathbf{k}}{\rm Im}[g_{mm}^{R}(\mathbf{k},\mathbf{k}+\mathbf{q},\omega)+g_{mm}^{R}(\mathbf{k},\mathbf{k}-\mathbf{q},\omega)-g_{m+3m+3}^{A}(\mathbf{k},\mathbf{k}+\mathbf{q},-\omega)-g_{m+3m+3}^{A}(\mathbf{k},\mathbf{k}-\mathbf{q},-\omega)]\nonumber\\
&&+i{\rm Re}[g_{mm}^{R}(\mathbf{k},\mathbf{k}+\mathbf{q},\omega)-g_{mm}^{R}(\mathbf{k},\mathbf{k}-\mathbf{q},\omega)-g_{m+3m+3}^{A}(\mathbf{k},\mathbf{k}+\mathbf{q},-\omega)+g_{m+3m+3}^{A}(\mathbf{k},\mathbf{k}-\mathbf{q},-\omega)].
\end{eqnarray}
\end{widetext}
In both the models, since the $d_{xy}$ orbital does not mix with the $d_{xz}/d_{yz}$ orbital and the impurity scattering we assumed is purely intraorbital, thus there cannot be quasiparticle scattering between the $\gamma$ and $\alpha/\beta$ bands. We then define the difference of $\rho(\mathbf{q},\omega)$ between the SC and normal states as
\begin{eqnarray}
\label{dif}
\Delta\rho(\mathbf{q},\omega)=\rho(\mathbf{q},\omega)|_{SC}-\rho(\mathbf{q},\omega)|_{normal}.
\end{eqnarray}
In this way, the contribution to the LDOS from $(\alpha,\beta)$ bands and from the $\gamma$ band can be completely disentangled. This provides a unique possibility to probe the active/passive bands.

\begin{figure}
\includegraphics[width=1\linewidth]{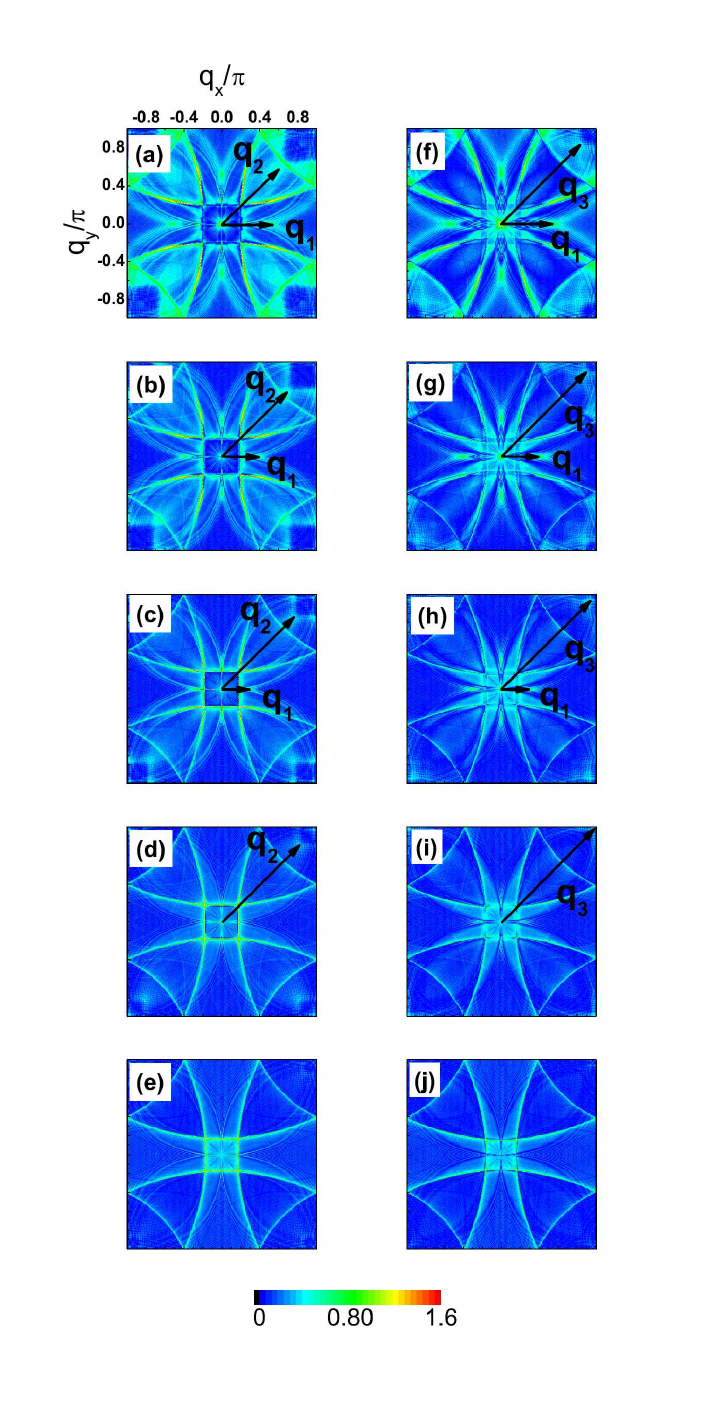}
 \caption{\label{2dqpi} (color online) $|\Delta\rho(\mathbf{q},\omega)|$ for the 2D model. The point at $\mathbf{q}=0$ is neglected in order to show weaker features at other wave vectors. From (a) to (e), $\omega/\Delta_{0}=-1,-0.8,-0.6,-0.4,-0.2$ while from (f) to (j), $\omega/\Delta_{0}=1,0.8,0.6,0.4,0.2$.}
\end{figure}

\begin{figure}
\includegraphics[width=1\linewidth]{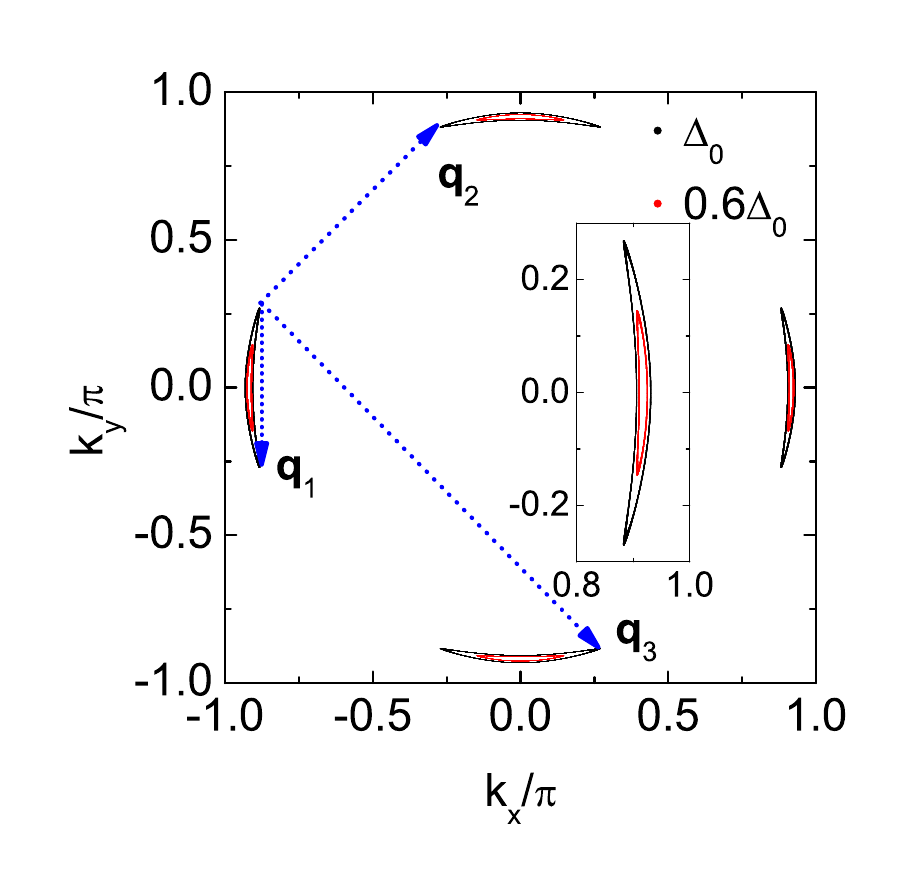}
 \caption{\label{2dcec} (color online) The banana-shaped CEC of the $\gamma$ band for the 2D model, at $|\omega|=\Delta_{0}$ (black) and $0.6\Delta_{0}$ (red). The tips of the CEC trace the normal state FS of the $\gamma$ band. The arrows indicate the three main scattering wave vectors. The inset shows the CEC around $(k_{x}/\pi,k_{y}/\pi)=(0.92,0)$.}
\end{figure}

\begin{figure}
\includegraphics[width=1\linewidth]{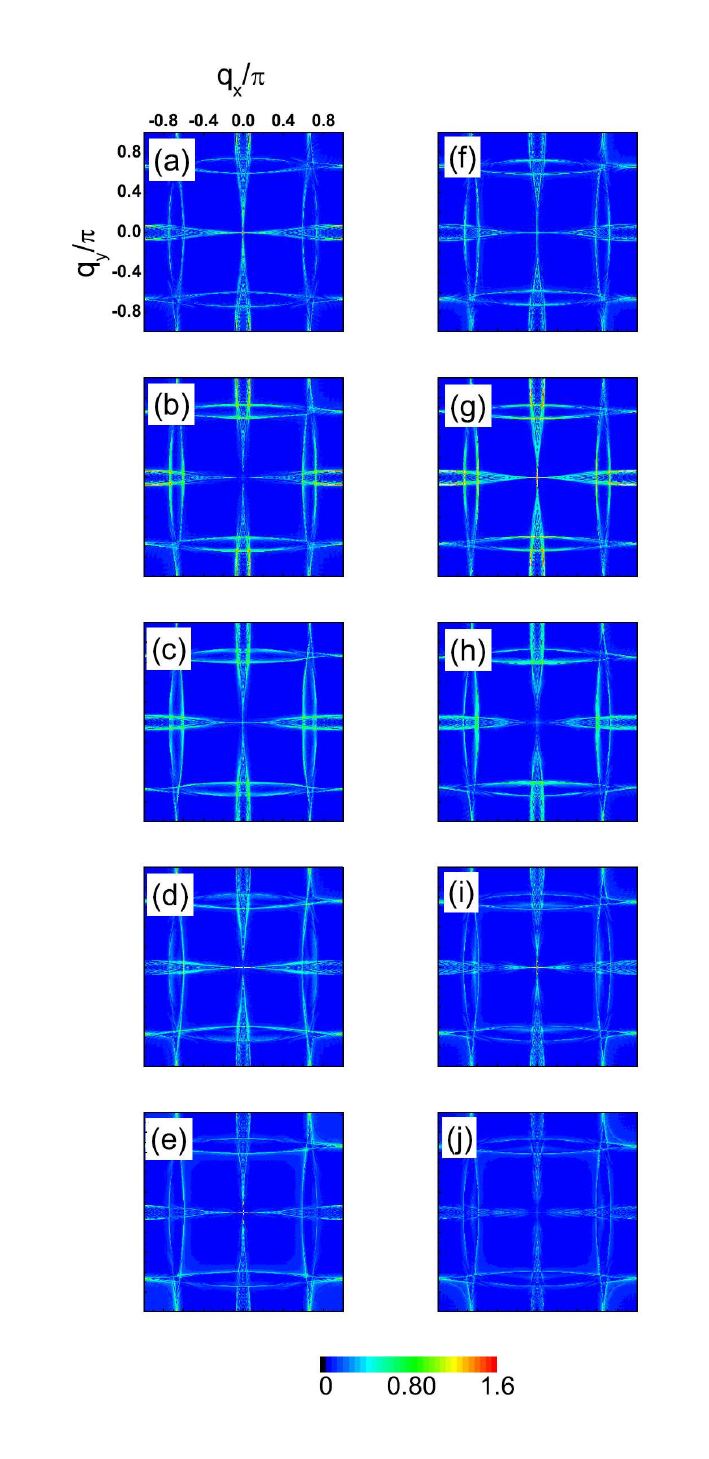}
 \caption{\label{1dqpi} (color online) The same as Fig. \ref{2dqpi}, but for the 1D model.}
\end{figure}

\begin{figure}
\includegraphics[width=1\linewidth]{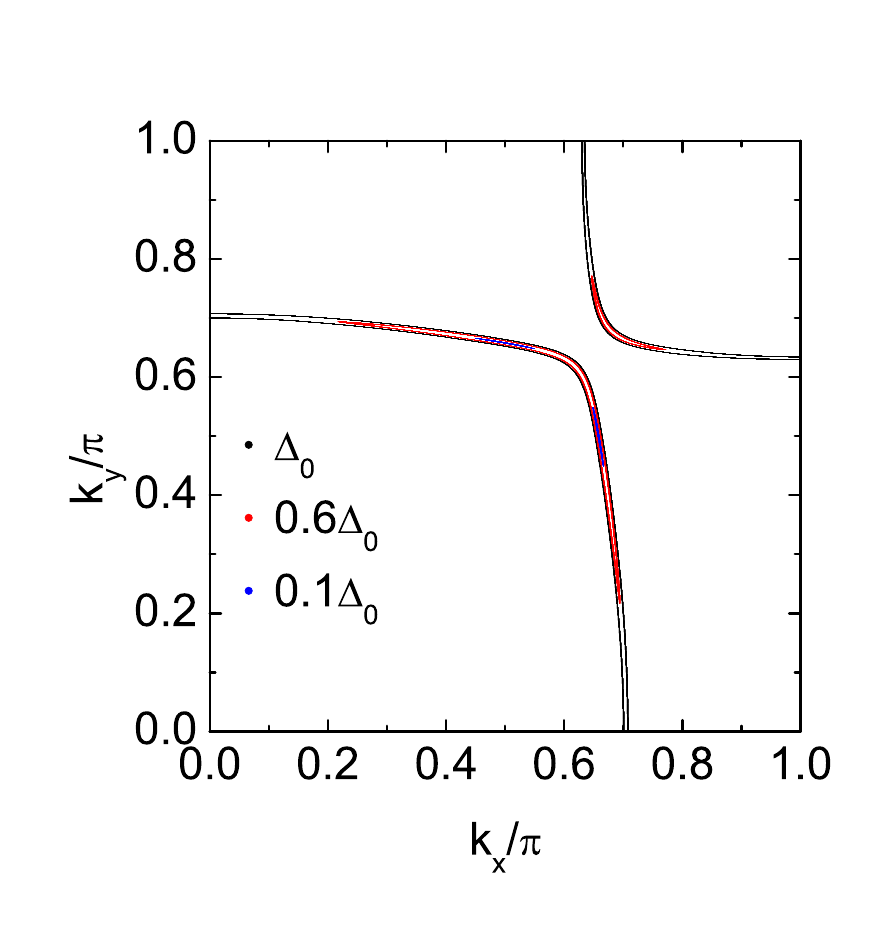}
 \caption{\label{1dcec} (color online) The CEC of the $\alpha$ and $\beta$ bands for the 1D model, at $|\omega|=\Delta_{0}$ (black), $0.6\Delta_{0}$ (red) and $0.1\Delta_{0}$ (blue). Here we show only the first quadrant of the BZ.}
\end{figure}

\emph{Results}.---For the 2D model, we plot $|\Delta\rho(\mathbf{q},\omega)|$ in Fig. \ref{2dqpi} and three main scattering wave vectors $\mathbf{q}_{1}$, $\mathbf{q}_{2}$ and $\mathbf{q}_{3}$ can be identified [see the arrows in Figs. \ref{2dqpi}(a) and \ref{2dqpi}(f)]. $\mathbf{q}_{1}$ is along the $q_{x}=0$ and $q_{y}=0$ directions while $\mathbf{q}_{2}$ and $\mathbf{q}_{3}$ are along the $q_{x}=\pm q_{y}$ directions. When $|\omega|$ decreases, $\mathbf{q}_{1}$ moves towards the origin while $\mathbf{q}_{2}$ and $\mathbf{q}_{3}$ move towards $(\pm\pi,\pm\pi)$. The $\omega$ dependence of $\mathbf{q}$ can be understood from the evolution of the constant-energy contour (CEC).~\cite{QPI,hoffman,zhu} In the normal state, the CEC of the three bands shows little variation when $|\omega|\leqslant\Delta_{0}$ since $\Delta_{0}\ll1$, thus in this energy range $\rho(\mathbf{q},\omega)$ barely evolves with $\omega$. On the contrary, in the SC state, first of all, the CEC of the $\alpha$ and $\beta$ bands is the same as that in the normal state since the SC pairing {is tiny and set to zero in these bands within our approximation}, thus the contribution to the QPI from these two bands can be completely removed from Eq. (\ref{dif}). For the $\gamma$ band, on the other hand, superconductivity gaps the entire Fermi surface (FS) with deep gap minima ($\approx0.14\Delta_{0}$) at $(k_{x}/\pi,k_{y}/\pi)\approx(\pm0.92,0)/(0,\pm0.92)$. From Fig. \ref{2dcec} we can see, around each of the four gap minima, the CEC evolves from a single point at the gap minima, to banana-shaped closed contour at higher energies. The size of the banana increases with $|\omega|$ and its tips trace the normal state FS. By carefully examining the locations of these banana tips, we conclude that the scattering wave vectors in Fig. \ref{2dqpi} should correspond to the wave vectors connecting the banana tips (see the blue dotted arrows in Fig. \ref{2dcec}). For example, in Fig. \ref{2dqpi}(a), at $\omega/\Delta_{0}=-1$, $\mathbf{q}_{1}/\pi\approx(\pm0.54,0)/(0,\pm0.54)$ and $\mathbf{q}_{2}/\pi\approx(\pm0.6,\pm0.6)$. At the same $\omega$, the CEC tips in Fig. \ref{2dcec} are located at $(\pm0.88,\pm0.27)\pi/(\pm0.27,\pm0.88)\pi$. The derived $\mathbf{q}_{1}/\pi$ and $\mathbf{q}_{2}/\pi$ are $(\pm0.54,0)/(0,\pm0.54)$ and $(\pm0.61,\pm0.61)$, respectively, agree fairly well with those in Fig. \ref{2dqpi}(a). As $\omega/\Delta_{0}$ changes from $-1$ to $-0.2$, the size of the banana decreases, thus from Fig. \ref{2dcec}, $|\mathbf{q}_{1}|$ should decrease and $|\mathbf{q}_{2}|$ should increase. As we can see from Figs. \ref{2dqpi}(a) to \ref{2dqpi}(e), the evolution of $\mathbf{q}_{1}$ and $\mathbf{q}_{2}$ indeed follows this trend. On the other hand, at $\omega/\Delta_{0}=1$, $\mathbf{q}_{3}$ in Fig. \ref{2dqpi}(f) is located at $(\pm0.84,\pm0.84)\pi$ and that derived from Fig. \ref{2dcec} is at $(\pm1.15,\pm1.15)\pi$. They differ by a reciprocal lattice constant, suggesting that $\mathbf{q}_{3}$ is an umklapp process. In Fig. \ref{2dcec}, as $|\omega|$ decreases, $|\mathbf{q}_{3}|$ should decrease and since it is an umklapp process, therefore $\mathbf{q}_{3}$ in Figs. \ref{2dqpi}(f) to \ref{2dqpi}(j) moves towards $(\pm\pi,\pm\pi)$. In addition we found that the spectra are asymmetrical with respect to positive and negative $\omega$. The asymmetry becomes more obvious as $|\omega|$ increases. Furthermore, $\mathbf{q}_{2}$ and $\mathbf{q}_{3}$ are invisible for positive and negative $\omega$, respectively.

For the 1D model, the SC pairing becomes more complicated. From Eq. (\ref{oned}) we can see that the pairing is purely intraorbital. However, since $\Delta_{\mathbf{k}}^{11}\neq\Delta_{\mathbf{k}}^{22}$, thus in band space, there exists interband pairing term such as $c_{\mathbf{k}\alpha\uparrow}^{\dag}c_{\mathbf{-k}\beta\downarrow}^{\dag}$ and the quasiparticle energies derived from the $d_{xz}$ and $d_{yz}$ orbitals can be written as
\begin{eqnarray}
\xi^{\pm}_{\mathbf{k}}&=&\frac{1}{\sqrt{2}}\Big{\{}\lambda_{\mathbf{k}}\pm\Big{[}\lambda^{2}_{\mathbf{k}}-4(\epsilon_{1\mathbf{k}}^{2}\epsilon_{2\mathbf{k}}^{2}-2\epsilon_{1\mathbf{k}}\epsilon_{2\mathbf{k}}|\epsilon_{12\mathbf{k}}|^{2}\nonumber\\
&&+|\epsilon_{12\mathbf{k}}|^{4}+\epsilon_{2\mathbf{k}}^{2}|\Delta_{\mathbf{k}}^{11}|^{2}+\epsilon_{1\mathbf{k}}^{2}|\Delta_{\mathbf{k}}^{22}|^{2}\nonumber\\
&&+|\epsilon_{12\mathbf{k}}|^{2}\Delta_{\mathbf{k}}^{11*}\Delta_{\mathbf{k}}^{22}+|\epsilon_{12\mathbf{k}}|^{2}\Delta_{\mathbf{k}}^{11}\Delta_{\mathbf{k}}^{22*}\nonumber\\
&&+|\Delta_{\mathbf{k}}^{11}\Delta_{\mathbf{k}}^{22}|^{2})\Big{]}^{\frac{1}{2}}\Big{\}}^{\frac{1}{2}},
\end{eqnarray}
where
\begin{eqnarray}
\lambda_{\mathbf{k}}&=&\epsilon_{1\mathbf{k}}^{2}+\epsilon_{2\mathbf{k}}^{2}+2|\epsilon_{12\mathbf{k}}|^{2}+|\Delta_{\mathbf{k}}^{11}|^{2}+|\Delta_{\mathbf{k}}^{22}|^{2}.
\end{eqnarray}
For the $\Delta_{0}$ chosen in this paper, the quasiparticle energy $\xi^{-}_{\mathbf{k}}$ is parametrically small at $(k_{x}/\pi,k_{y}/\pi)\approx(\pm0.66,\pm0.50)/(\pm0.50,\pm0.66)$, which is about $0.02\Delta_{0}$. In Fig. \ref{1dqpi} we plot $|\Delta\rho(\mathbf{q},\omega)|$ for this model and in this case, the contribution to the LDOS from the $\gamma$ band is removed since there is no superconductivity in this band. We found, (1) the spectra resemble the shape of the $\alpha/\beta$ FS and are also quasi-one-dimensional, with additional features along the $q_{x}=0$ and $q_{y}=0$ directions. (2) There are no clear scattering wave vectors evolving with energy, thus the QPI signal is almost dispersionless. (3) The spectra show minor asymmetry with respect to positive and negative $\omega$ as compared to those of the 2D model shown in Fig. \ref{2dqpi}. Most importantly, the QPI in this model cannot be described by scatterings on the CEC alone. As we can see from Fig. \ref{1dcec}, in the first quadrant of the Brillouin zone (BZ), the CEC evolves from two separate points at $(k_{x}/\pi,k_{y}/\pi)\approx(0.66,0.50)/(0.50,0.66)$ at $|\omega|\approx0.02\Delta_{0}$, to closed contours at higher energies and the tips of the CEC trace the original FS of the $\alpha$ and $\beta$ bands. However in Fig. \ref{1dqpi} we cannot find any scattering wave vectors associated with the evolution of the CEC in Fig. \ref{1dcec}, implying that the off-shell contributions to the the scatterings beyond the CEC become more important. The reason of this phenomenon may be: Due to the existence of the interband pairing terms, the CEC in the SC state mixes the $\alpha$ and $\beta$ bands at different energies, therefore those off-shell scatterings may also contribute to the QPI significantly.

\emph{Summary}.---In summary, we have studied the QPI in Sr$_{2}$RuO$_{4}$ based on two different pairing models in order to propose an experimental method to test them. For the 2D model, the QPI spectra are two-dimensional, with clear peaks evolving with energy and their locations can be determined from the tips of the CEC. On the contrary, for the 1D model, the QPI spectra are quasi-one-dimensional and almost dispersionless, which may involve off-shell contributions to the scatterings beyond the CEC. Since the QPI can be directly measured by STM, therefore the distinct differences of the QPI between these two models can help to resolve the controversy of in which bands superconductivity develops. In addition, for both the models, we repeated the above calculations for $V_{s}=8\Delta_{0}$ and the results remain qualitatively the same, indicating the robustness of the QPI spectra presented in this paper.

This work was supported by NSFC (Grants No. 11204138 , No. 11175087 and No.11023002), the Ministry of Science and Technology of China (Grants No. 2011CBA00108 and 2011CB922101), NSF of Jiangsu Province of China (Grant No. BK2012450), Program of Natural Science Research of Jiangsu Higher Education Institutions of China (Grant No. 12KJB140009), SRFDP (Grant No. 20123207120005), China Postdoctoral Science Foundation (Grant No. 2012M511297), NCET (Grant No. NCET-12-0626), the Texas Center for Superconductivity and the Robert A. Welch Foundation under Grant No. E-1146.


\begin{thebibliography}{99}

\bibitem{maeno} Y. Maeno, H. Hashimoto, K. Yoshida, S. Nishizaki, T. Fujita, J. G. Bednorz, and F. Lichtenberg,  Nature \textbf{372}, 532 (1994).

\bibitem{rice} T. M. Rice and M. Sigrist, J. Phys.: Condens. Matter \textbf{7}, L643 (1995).

\bibitem{baskaran} G. Baskaran, Physica B \textbf{223}-\textbf{224}, 490 (1996).

\bibitem{nelson} K. D. Nelson, Z. Q. Mao, Y. Maeno, and Y. Liu, Science \textbf{306}, 1151 (2004).

\bibitem{duffy} J. A. Duffy, S. M. Hayden, Y. Maeno, Z. Mao, J. Kulda, and G. J. McIntyre, Phys. Rev. Lett. \textbf{85}, 5412 (2000).

\bibitem{ishida2} K. Ishida, H. Mukuda, Y. Kitaoka, K. Asayama, Z. Q. Mao, Y. Mori, and Y. Maeno, Nature \textbf{396}, 658 (1998).

\bibitem{luke} G. M. Luke, Y. Fudamoto, K. M. Kojima, M. I. Larkin, J. Merrin, B. Nachumi, Y. J. Uemura, Y. Maeno, Z. Q. Mao, Y. Mori, H. Nakamura, and M. Sigrist, Nature \textbf{394}, 558 (1998).

\bibitem{kidwingira} F. Kidwingira, J. D. Strand, D. J. Van Harlingen, and Y. Maeno, Science \textbf{314}, 1267 (2006).

\bibitem{xia} J. Xia, Y. Maeno, P. T. Beyersdorf, M. M. Fejer, and A. Kapitulnik, Phys. Rev. Lett. \textbf{97}, 167002 (2006).

\bibitem{mackenzie} A. P. Mackenzie and Y. Maeno, Rev. Mod. Phys. \textbf{75}, 657 (2003).

\bibitem{kallin} C. Kallin, Rep. Prog. Phys. \textbf{75}, 042501 (2012).

\bibitem{read} N. Read and D. Green, Phys. Rev. B \textbf{61}, 10267 (2000).

\bibitem{nayak} C. Nayak, S. H. Simon, A. Stern, M. Freedman, and S. D. Sarma, Rev. Mod. Phys. \textbf{80}, 1083 (2008).

\bibitem{oguchi} T. Oguchi, Phys. Rev. B \textbf{51}, 1385 (1995).

\bibitem{mackenzie2} A. P. Mackenzie, S. R. Julian, A. J. Diver, G. J. McMullan, M. P. Ray, G. G. Lonzarich, Y. Maeno, S. Nishizaki, and T. Fujita, Phys. Rev. Lett. \textbf{76}, 3786 (1996).

\bibitem{shen} A. Damascelli, D. H. Lu, K. M. Shen, N. P. Armitage, F. Ronning, D. L. Feng, C. Kim, Z.-X. Shen, T. Kimura, Y. Tokura, Z. Q. Mao, and Y. Maeno, Phys. Rev. Lett. \textbf{85}, 5194 (2000).

\bibitem{deguchi} K. Deguchi, Z. Q. Mao, H. Yaguchi, and Y. Maeno, Phys. Rev. Lett. \textbf{92}, 047002 (2004).

\bibitem{Ishida} K. Ishida, H. Mukuda, Y. Kitaoka, Z. Q. Mao, Y. Mori, and Y. Maeno, Phys. Rev. Lett. \textbf{84}, 5387 (2000).

\bibitem{Agterberg} D. F. Agterberg, T. M. Rice, and M. Sigrist, Phys. Rev. Lett. \textbf{78}, 3374 (1997).

\bibitem{miyake} K. Miyake and O. Narikiyo, Phys. Rev. Lett. \textbf{83}, 1423 (1999).

\bibitem{ng} K. K. Ng and M. Sigrist, Europhys. Lett. \textbf{49}, 473 (2000).

\bibitem{kirtley1} J. R. Kirtley, C. Kallin, C. W. Hicks, E.-A. Kim, Y. Liu, K. A. Moler, Y. Maeno, and K. D. Nelson, Phys. Rev. B {\bf 76}, 014526 (2007).

\bibitem{kirtley2} C. W. Hicks, J. R. Kirtley, T. M. Lippman, N. C. Koshnick, M. E. Huber, Y. Maeno, W. M. Yuhasz, M. B. Maple, and K. A. Moler, Phys. Rev. B {\bf 81}, 214501 (2010).

\bibitem{raghu} S. Raghu, A. Kapitulnik, and S. A. Kivelson, Phys. Rev. Lett. \textbf{105}, 136401 (2010).

\bibitem{wang} Q. H. Wang, C. Platt, Y. Yang, C. Honerkamp, F. C. Zhang, W. Hanke, T. M. Rice, and R. Thomale, arXiv:1305.2317.

\bibitem{QPI} Q. H. Wang and D. H. Lee, Phys. Rev. B {\bf 67}, 020511 (2003).

\bibitem{hoffman} J. E. Hoffman, K. McElroy, D.-H. Lee, K. M Lang, H. Eisaki, S. Uchida, and J. C. Davis, Science \textbf{297}, 1148 (2002).

\bibitem{zhu} A. V. Balatsky, I. Vekhter, and Jian-Xin Zhu, Rev. Mod. Phys. \textbf{78}, 373 (2006).



\end{thebibliography}
\end{document}